\documentclass[a4paper,12pt]{article}
\newcommand{\be}{\begin{equation}}
\newcommand{\ee}{\end{equation}}
\newcommand{\ben}{\begin{eqnarray}}
\newcommand{\een}{\end{eqnarray}}
\newcommand{\tgvv}{\tilde{g}_{w\bar{w}}}
\newcommand{\tgss}{\tilde{g}_{t\bar{t}}}
\newcommand{\tgvs}{\tilde{g}_{w\bar{t}}}
\newcommand{\ls}{{l_s}}
\newcommand{\del}{\partial}
\newcommand{\vp}{\varphi}
\newcommand{\non}{\nonumber}
\newcommand{\ba}{\begin{array}}
\newcommand{\ea}{\end{array}}
\newcommand{\R}{{\bf R}}
\newcommand{\journal}[4]{{\rm #1} {\bf #2} (19#3) #4}
\newcommand{\NP}{\journal{Nucl. Phys.}}
\newcommand{\PL}{\journal{Phys. Lett.}}
\newcommand{\PRL}{\journal{Phys. Rev. Lett.}}

\newcommand{\JHEP}{\journal{JHEP}}

\newcommand{\ATMP}{\journal{Adv.Theor.Math.Phys.}}
\makeatletter
 
 \@addtoreset{equation}{section}
\makeatother
\begin{document}

\begin{titlepage}
\begin{flushright}
       {\normalsize  KUNS-1701 \\
       hep-th/0012030  \\
           December 2000}
\end{flushright}
\vspace{10mm}
\begin{center}
  {\Large \bf Linear Dilaton Background \\  and \\ \vspace{3mm} Fully Localized
Intersecting Five-branes}

\renewcommand{\thefootnote}{\fnsymbol{footnote}}
\vfill
{ Kazutoshi ~Ohta}\footnote{e-mail address: kohta@yukawa.kyoto-u.ac.jp}
and
{ Takashi ~Yokono}\footnote{e-mail address: yokono@gauge.scphys.kyoto-u.ac.jp}\\
         \vspace{2.5mm}
   {\em     Department of Physics, Kyoto University, Kyoto 606-8502, Japan}\\
\renewcommand{\thefootnote}{\arabic{footnote}}
\end{center}
\vfill
\begin{abstract}
We investigate a near-horizon geometry of NS5-branes wrapping on a
Riemann surface, which asymptotically approaches to linear dilaton
backgrounds. We concretely find a fully localized solution of the
near-horizon geometry of intersecting NS5-branes. We also discuss a
relation to a description of Landau-Ginzburg theories.
\end{abstract}
\vfill
\end{titlepage}

\newpage

\section{Introduction}

Non-gravitational theories in six dimensions having string-like
properties, which are called ``little string theories'' (LSTs), have
been attracted much interests. (See for review \cite{Aharony}.) There is
no consistent Lagrangian and these theories are considered as non-local
field theories.

To understand the LSTs we have various approaches from superstring
theories, even though there is no field theoretical treatment. The first
construction of the LSTs \cite{Seiberg} was obtained by using discrete light-cone
quantization (DLCQ) of M(atrix) theory \cite{BFSS}. More intuitive
definition is world-volume effective theory of $N$ parallel NS5-branes
in the limit taking the string coupling to zero ($g_s\rightarrow 0$) and
fixing the string scale. The LSTs can be holographically understood from
the near-horizon geometry of parallel NS5-branes \cite{ABKS}. This
approach is T-dual equivalent to Type II string theory on an $A_{k-1}$
singularity with $g_s\rightarrow 0$ \cite{OV, Kutasov}.

Originally, LSTs are six dimensional
non-gravitational theories, but we expect that there exist lower
dimensional LSTs by compactifying the 6d LSTs. For example, if we
consider NS5-branes whose world-volume is $\R^{1,3}\times \Sigma$, where
$\Sigma$ is a compact two-dimensional Riemann surface, we obtain a four
dimensional LST. When we especially choose the Riemann surface as a
Seiberg-Witten curve and lift the NS5-brane to M5-brane wrapping on
$\Sigma$ in M-theory \cite{Witten}, we expect a relation between some
phase of four dimensional QCD and the LST.

Holographic analysis for this four dimensional LST needs a near-horizon
of SUGRA solution of the NS5-branes wrapping on the $\Sigma$. 
Some partially and fully localized solutions were constructed by \cite{Youm, FS1, AR}. 
The fully localized near-horizon geometry wrapping on the $\Sigma$, 
which is the
warped AdS geometry, was found by \cite{FS2}. However, we now need some
linear dilaton backgrounds rather than the AdS in order to investigate
the four dimensional LSTs which are not conformal field theory. In this
paper, we search for the near-horizon SUGRA solution of the NS5-brane
wrapping on the $\Sigma$ as a kind of Fayyazuddin-Smith geometry
\cite{FS1}. Especially, we find the solution in the case of highly
degenerate Riemann surface, which describes intersecting NS5-branes with
four dimensional common world-volume.

Our solution have very similar structure to the parallel NS5-branes
solution by using a suitable singular coordinate
transformation. Superstring theory on the near-horizon geometry of the
$N$ parallel NS5-branes has a dual description to a Landau-Ginzburg (LG)
theory with a suitable superpotential, which have the same form as a
curve for ALE space with $A_{N-1}$ singularity \cite{OV, Kutasov}. If we
apply these results to our solution, we can expect that dual theory on
the NS5-branes intersection relates to the LG theory with a
superpotential of a curve for a singular Calabi-Yau three fold. This
result agrees with \cite{GKP}.

This paper is organized as follows. We start with a brief review of the
Fayyazuddin-Smith geometry and discuss some decoupling limits of this
geometry in a subsequent section. In section \ref{secLD} we construct a fully
localized solution of the intersecting NS5-branes which enjoys a linear
dilaton form. We assume the linear dilation form at first in the
approach to the solution in this section. Alternative general
construction, in which we first make a intersection M5-branes solution
in M-theory, is given in section \ref{secGSOL}. These two approaches are consistent
with each other. In section \ref{secLG} we will discuss a relation to the LG
model. The final section is devoted to conclusion and discussion.

\section{A brief review of the Fayyazuddin-Smith geometry}
In this section, we review some results on the M5-brane solution wrapping on a Riemann surface \cite{FS1}.

The general solution for M5-brane configurations which preserve at least 1/4 
supersymmetries is studied in \cite{FS1}.  Such M5-branes have world-volume 
directions along $x^0,x^1,x^2,x^3 $ and wrap on a Riemann surface holomorphically
embedded in ${\bf C}^2 \ni (v,s)$, where we define $v\equiv x^4+ ix^5$ and $s=x^6 + i x^7$. The overall transverse directions to the M5-branes are $x^8,x^9,x^{10}$.
The metric is given in \cite{FS1}, 
\be
ds^2 = H^{-\frac{1}{3}} dx_{3+1}^2 + H^{-\frac{1}{3}} g_{m\bar{n}} dz^m dz^{\bar{n}} + H^{\frac{2}{3}} \delta_{\alpha\beta} dx^{\alpha} dx^{\beta}, \label{FSoriginal}
\ee
where $\alpha, \beta = 8,9,10$, $m,n=v,s$ and $g_{m\bar{n}}$ is required to be a K\"ahler metric. Here $H$ is the determinant of $g_{m\bar{n}}$,
\be
H = g_{v\bar{v}} g_{s\bar{s}} - g_{v\bar{s}} g_{s\bar{v}}.
\ee
The source equation is 
\be
{\it d} F = J_{m\bar{n}} dz^{m} \wedge dz^{\bar{n}} \wedge dx^8 \wedge dx^9 \wedge dx^{10}
\ee
where $F$ is 4-form field strength and 
\be
J_{m\bar{n}} = \del_{\alpha} \del_{\alpha} g_{m\bar{n}} + 4 \del_m \del_{\bar{n}} H. \label{SOURCE}
\ee
The source $J$  specifies the position of M5-branes. For example, the source for the intersecting N M5-branes whose world-volume are the $x^0, x^1, x^2, x^3, x^4,x^5$ directions and N$'$ M5$'$-branes whose world-volume are the $x^0, x^1, x^2, x^3, x^6,x^7$ directions is given by
\ben
J_{s\bar{s}} &=& N\, l_p^3\, \delta^{(3)}(x^{\alpha}) \,\delta^{(2)}(s), \\
J_{v\bar{v}} &=& N'\, l_p^3\, \delta^{(3)}(x^{\alpha})\,  \delta^{(2)}(v),\\
J_{v\bar{s}} &(=& J_{s\bar{v}}) = 0. \label{SOURCEend}
\een

Given a M5-brane configuration which determine the source equation
(\ref{SOURCE}), solving it and K\"ahlar condition on $g_{m\bar{n}}$, we
have the SUGRA solution for the M5-brane configuration. In other words,
if we have the metric which satisfies the K\"ahler condition, then we
find the M5-brane configuration by using the source equation
(\ref{SOURCE}).

\section{Decoupling limit}
The world-volume theory on NS5-branes was studied in the decoupling limit $g_s \to 0$  with $\ls $ fixed \cite{Seiberg}, where $g_s $ is the string coupling 
constant and $\ls $ is the string length. In this limit, the modes in the bulk of space-time decouple, while 
the dynamics of modes on the NS5-branes remains. 

Since NS5-branes are dual to the M5-branes which are localized on the
eleventh (compactified) direction, we consider the decoupling limit in
the M-theory. The parallel 5-branes case is studied in \cite{ABKS} and
the six-dimensional decoupled theory appears. In the configurations
which we will consider in this paper, there are common four-directions
on the 5-branes. Hence we will have the four-dimensional decoupled
theory.  The string length $\ls $ of Type IIA theory is given by $ \ls^2
= l_p^3/ R$, where $R$ is the radius of the compactified direction. We
take the limit in which $R$ and $l_p $ go to zero with $\ls$ fixed.

We choose the natural coordinates in this limit as follows. We would
like to keep the excitations which remain interacting on the M5-branes
in this limit.  The M2-branes may end on the M5-branes. On the M5-brane
world-volume theory, two types of BPS objects appear. We would like to
fix the their mass. The first is that mass of the M2-branes which
streched between the M5-branes along the overall transverse directions
$x^{\alpha}$ appears as $x^{\alpha}/l_p^3$. The other is M2-branes which
have a disc topology bounded on a cycle of the Riemann surface in
$(v,s)$-space.  The mass of the M2-branes is given by the area of the
disk and appears as $vs/l_p^3$.  Thus we choose the natural coordinates
as
\ben
y^{\alpha} &=& \frac{x^{\alpha}}{l_p^3}, \\
w\, t &=& \frac{v\, s}{l_p^3}.
\een
In terms of $y, w, t$, the M5-branes metric (\ref{FSoriginal}) becomes
\be
\frac{ds^2}{l_p^2} = H^{-\frac{1}{3}} dx_{3+1}^2 + H^{-\frac{1}{3}} g_{m\bar{n}} dz^m dz^{\bar{n}} + H^{\frac{2}{3}} \delta_{\alpha\beta} dy^{\alpha} dy^{\beta},\label{FSnew}
\ee
where $m,n$ represent $w,t$ and we redefine as $H = g_{w\bar{w}} g_{t\bar{t}} - g_{w\bar{t}} g_{t\bar{w}}$. The $l_p^3$ factors in the source equations disappear. For example, the source equations for the intersecting M5-branes (\ref{SOURCE})-(\ref{SOURCEend}) become
\ben
\del_{\alpha} \del_{\alpha} g_{t\bar{t}} + 4 \del_t \del_{\bar{t}} H &=&
N\,  \delta^{(3)} (y^{\alpha})\, \delta^{(2)}(t), \\
\del_{\alpha} \del_{\alpha} g_{w\bar{w}} + 4 \del_w \del_{\bar{w}} H &=&
N'\,  \delta^{(3)} (y^{\alpha})\, \delta^{(2)}(w), \\
\del_{\alpha} \del_{\alpha} g_{w\bar{t}} + 4 \del_w \del_{\bar{t}} H &=&
0.
\een
Since the source equations do not have $l_p$ factor, solving these equations, 
the solutions of $g_{m\bar{n}}$  do not include $l_p$ factor. 
Therefore the metric (\ref{FSnew}) does not depend on $l_p$ except for the 
overall $l_p$ factor. This $l_p$ factor will not effect on any physical 
computations as in the case of a similar factor in \cite{Maldacena}.

The assignment of the $l_p$ factor between $v$ and $s$ does not change above results. But we would like to treat $v$ and $s$ equally, we choose as
\ben
w &=& \frac{v}{l_p^{\frac{3}{2}}}, \\
t &=& \frac{s}{l_p^{\frac{3}{2}}}.
\een

In the next section, we will search for the 5-brane solution in the decoupling limit.

\section{Linear dilaton solution} \label{secLD}
It was proposed \cite{ABKS} that any vacua of string theories which have
asymptotic linear dilaton backgrounds are holographic. The lower
dimensional decoupled theories, which are in general not local QFT, are
described by string theory on the backgrounds,
\ben ds^2 &=& dx_{1,d-1}^2 + d\phi^2 + ds^2({\cal M}), \label{LDF}\\
g_s^2 &=& e^{-Q\phi},
\een where $Q$ is a constant and $\phi $ relates to the
dilaton linearly. We call this $\phi$ the dilation in the following. The
metric on the $9-d$ dimensional manifold ${\cal M}$ is independent of
$x$ and $\phi $.

Authors in \cite{ABKS} commented that NS5-branes wrapping on a Riemann
surface which is considered in this paper is a possible candidate for
the proposition, but the linear dilaton background for such
configuration has been not known.

Another approach to the decoupled theory was discussed in \cite{GKP,GK}. 
String theory on $\R^{1,d-1} \times X^{2n}$ was studied
in the decoupling limit. 
Here $2n= 10-d$ and $X^{2n}$ is a singular (non-compact) Calabi-Yau $n-$fold ($CY_n$).  
The decoupled theory arises near the singularity on $X^{2n}$, and
 is dual to 
string theory on the linear background which is
\be
\R^{1,d-1} \times \R_{\phi} \times {\cal N}
\ee
where ${\cal N} $ is the manifold at a fixed distance from the singular point on $X^{2n}$. 

For $n=3$ case, non-compact $CY_3$ is considered as T-dual to the
NS5-branes wrapping a Riemann surface \cite{Vafa}. Since string theory
on $(CY_n)$ is dual to string theory on the linear dilaton background,
the metric for NS5-branes wrapping a Riemann surface expect to be also
the linear dilaton form. Thus we look for the NS5-brane solutions which
have the linear dilaton background.

We have the type IIA NS5-brane solution by the dimensional reduction
along one of the overall transverse directions to the M5-brane solution
(\ref{FSnew}). The standard relation between 11-dimensional supergravity
and type IIA supergravity is given in \cite{Witten2, BHO}. We compactify
along the eleventh direction. The NS5-brane metric becomes
\be
ds^2 = dx_{3+1}^2 + g_{m\bar{n}} dz^m d\bar{z}^{\bar{n}} + H dz d\bar{z}, \label{FSiia}
\ee
where we define $z= y^8 + i y^9$. The dilaton $\phi $ is determined by the compactification as $e^{\phi/\sqrt{N}\ls} = H^{1/2}$ where $N$ is a dimensionless real number. Thus, if the metric has the linear dilaton form (\ref{LDF}), it  might become
\be
ds^2= dx_{3+1}^2 + N \ls^2 \frac{ dH^2}{4 H^2} + ds^2  ({\cal M}). \label{LDF2}
\ee
We first seek the solutions such that the general solution (\ref{FSiia}) 
has the linear dilaton form (\ref{LDF2}). Then, we check if
the solutions satisfy the K\"ahler condition and the source equations.

Now we have three dimensionful complex coordinates $w,t$ and $z$ whose dimensions in length are $-1/2, -1/2$ and $-2$. 
It is convenient to choose only one dimensionful (radial) coordinate.
We define that 
\ben
w &=& \tilde{w}^a, \\
t &=& \tilde{t}^a, \\
z &=& \tilde{z}^b,
\een
where $a,b$ are some rational parameters.
Since we have the fixed string length $\ls$ as another dimensionful parameter,
 we select the dimensionful radial coordinate as 
\be
\rho^2 = \frac{ |\tilde{v}|^2 + |\tilde{s}|^2}{\ls^{2 c}}+\frac{|\tilde{z}|^2}{\ls^{2 d}}, \label{radius}
\ee
where $c,d$ are also rational parameters. We adopt the radial coordinate in terms of $\tilde{w},\tilde{t}, \tilde{z}$, not $w,t,z$, because the powers of original coordinates in the radial coordinate   often deviate from  usual ones in the case of 
localized solutions \cite{Youm, FS1, FS2, ITY}. 
The first term and second term in (\ref{radius}) must have the same dimension, so that
\be
\frac{1}{2 a} + c = \frac{2}{b} + d.
\ee
The radial coordinate $\rho $ is related to the dilaton $\phi $, and ${\bf R}^{1,3}$ and the manifold ${\cal M}$ does not depend on $\phi $. Therefore, since the manifold ${\cal M}$ does not depend on $\rho $ either, all elements related to $\rho $  in (\ref{FSiia}) come from the second term in (\ref{LDF2}) which have the $\ls^2$ factor.  Those in (\ref{FSiia}) must also have the $\ls^2$ factor, so that
\be
a c = b d = 1.
\ee
From these equations, we find that 
\be
b = 2a, \quad c = 1/a, \quad d = 1/2a.
\ee
So we parameterize $w,t,z$ as like follows,
\ben
w &=& \tilde{w}^a, \label{para1}\\
t &=& \tilde{t}^a,\\
z &=& \tilde{z}^{2 a},\\
\rho^2 &=& \frac{ |\tilde{w}|^2 + |\tilde{t}|^2}{\ls^{2/a}}+\frac{|\tilde{z}|^2}{\ls^{1/a}},\\
\tilde{w} &=& {l_s}^{\frac1a} \rho \sin\theta_1 \cos\theta_2 e^{i \varphi_1}, \\
\tilde{t} &=& {l_s}^{\frac1a} \rho \sin\theta_1 \sin\theta_2 e^{i \varphi_2}, \\
\tilde{z} &=& \ls^{\frac{1}{2a}} \rho \cos\theta_1 e^{i \varphi_3}. \label{para2}
\een
Here  $\theta_i\,(i=1,2)$ and $\vp_j \,(j=1,2,3)$ are dimensionless angular variables.

We now define that 
\ben
g_{w\bar{w}} &=& \rho^{- 2 a} \tgvv , \\
g_{t\bar{t}} &=& \rho^{-2 a} \tgss , \\
g_{w\bar{t}} &=& \rho^{-2 a} e^{ -a i (\vp_1-\vp_2)} \tgvs , \\
g_{t\bar{w}} &=& \rho^{-2 a} e^{ a i (\vp_1 - \vp_2)} \tgvs,
\een
and
\be
\tilde{H} \equiv  \tgvv \tgss - \tgvs^2 \left( = \rho^{4 a} H \right) . 
\ee
Here we assume that $\tgvv, \tgss$ and $\tgvs$ do not depend on the dimensionless combination $\rho \ls^\frac{3}{2a}$ , therefore $\tgvv , \tgss $ and $\tgvs $ are real functions which  depend only on  $\theta_1$ and $\theta_2$.
\footnote{
If we assume a K\"ahler potential $K=K(|w|, |t|, |z|)$, since $g_{m\bar{n}} = \del_m \del_{\bar{n}} K$ for $m,n=w,t$, $\tgvv, \tgss$ and $\tgvs$ do not depend on $\vp_j$.}   
By the definition of the parameterization (\ref{para1})-(\ref{para2}), we have the  overall $\ls^2$ factor in the general solution (\ref{FSiia}) except for $dx_{3+1}^2$ part. From the dimensional analysis, the metric elements $\tilde{g}_{kl}, k,l=\theta_i, \vp_j$  do not depend on $\rho $ because of above assumption. Since the manifold ${\cal M}$ does not depend on the dilaton (and $\rho $), all elements in (\ref{FSiia}) which relate to  $\rho$ come from the second term in the RHS of (\ref{LDF2}).

In these parameterization, the elements related to $\rho $ in the linear dilaton metric (\ref{LDF2}) are
\ben
g_{\rho\rho} &=& \frac{4 a^2 N \ls^2}{\rho^{2}}, \label{starteq}\\
g_{\rho\theta_1} &=& -\frac{2 a N \ls^2 \del_{\theta_1} \tilde{H}}{\rho}, \\
g_{\rho\theta_2} &=& - \frac{2 a N \ls^2 \del_{\theta_2} \tilde{H}}{\rho},\\
g_{\rho\vp_j} &=& 0 \quad \mbox{for}\quad j=1,2,3.
\een
On the other hand, those in the general solution (\ref{FSiia}) 
\ben
g_{\rho\rho} &=& \frac{a^2 \ls^2}{\rho^2} \left( 4 \cos^{4a} \theta_1 \tilde{H} + \sin^{2a}\theta_1 A \right) , \\
g_{\rho\theta_1} &=& - \frac{2 a^2 \ls^2}{\rho } \left(
 4 \cos^{4a-1} \theta_1 \sin\theta_1 \tilde{H} - \cos\theta_1 \sin^{2a-1}\theta_1 A \right) , \\
g_{\rho\theta_2} &=& -\frac{2 a^2 \ls^2}{\rho} \sin^{2a}\theta_1 \left(
\tgvv \cos^{2a-1}\theta_2 \sin\theta_2  - \tgss \cos\theta_2 \sin^{2a-1}\theta_2\right.  \non \\
&& \left. \qquad \qquad + \tgvs \left( \cos^{a-1}\theta_2 \sin^{a+1}\theta_2 - \cos^{a+1}\theta_2 \sin^{a-1}\theta_2 \right) \right), \\ \label{endeq}
g_{\rho\vp_j} &=& 0 \quad \mbox{for}\quad j=1,2,3.
\een
Here we define
\be
A \equiv  \left( \tgvv \cos^{2a} \theta_2 + 2 \tgvs \cos^a \theta_2 \sin^a \theta_2 + \tgss \sin^{2a}\theta_2 \right).
\ee
If the general solution become the linear dilaton form, the correspond elements in each metric must be equal.  Solving these equations (\ref{starteq})-(\ref{endeq}), we find
\be
\tilde{H} = \frac{N}{\cos^{4a}\theta_1 + f(\theta_2) \sin^{4a}\theta_1},
\ee
\ben
\tgvv &=& \frac{1}{B\cos^{2a} \theta_2\,
     \sin^{2a}\theta_1\,
      }
 \left( 64{a^2}N{{f(\theta_2)}^2}{\cos^4\theta_2}\,
      \sin^{4a}\theta_1 
  + 4\,{a^2}\,\cos^{4a}\theta_1\,
      \cos^{2a} \theta_2\,
      \sin^{2a} \theta_2 \right. \non \\
&&\hspace{-2cm}\left. + N{{f'(\theta_2)}^2}  \sin^{4a}\theta_1\,
      \sin^2 2 \theta_2
 + 4\,a\,f(\theta_2)\,
      \sin^{4a}\theta_1\,
      \left( a\,
         \cos^{2a} \theta_2\,
         \sin^{2a} \theta_2 
 - 8Nf'(\theta_2) \,\cos^3 \theta_2\,
         \sin \theta_2
         \right) \right) , \\
\tgss &=& \frac{1}{B
     \sin^{2a}\theta_1\,
     \sin^{2a} \theta_2}
 \left(64{a^2}N{{f(\theta_2)}^2}\,
      \sin^{4a}\theta_1\,
      {{\sin \theta_2}^4} 
 + 4\,{a^2}\,\cos^{4a}\theta_1\,
      \cos^{2a} \theta_2\,
      \sin^{2a} \theta_2  \right. \non \\
&&\hspace{-2cm}\left. +  N{{f'(\theta_2)}^2} \sin^{4a}\theta_1\,
      \sin^2 2 \theta_2
 + 4\,a\,f(\theta_2)\,
      \sin^{4a}\theta_1\,
      \left( a\,
         \cos^{2a} \theta_2\,
         \sin^{2a} \theta_2 + 
        8Nf'(\theta_2)\cos \theta_2\,
         \sin^3\theta_2
         \right) \right) , \\
\tgvs &=& \frac{-1}{B\cos^a\theta_2\,
     \sin^{2a}\theta_1\,
     \sin^a \theta_2}
\left( 4\,{a^2}\,
        \cos^{4a}\theta_1\,
        \cos^{2a} \theta_2\,
        \sin^{2a} \theta_2 
 - 16{a^2}N{{f(\theta_2)}^2}\,
        \sin^{4a}\theta_1\,
        \sin^2 2 \theta_2 \right. \non \\
&&\hspace{-1.5cm}\left. + N{{f'(\theta_2)}^2} \sin^{4a}\theta_1\,
        \sin^2 2 \theta_2\,
 + 4\,a\,f(\theta_2)\,
        \sin^{4a}\theta_1\,
        \left( a\,
           \cos^{2a} \theta_2\,
           \sin^{2a} \theta_2 - N
          f'(\theta_2)  \sin 4\,\theta_2
          \right)  \right) ,
\een
where $f(\theta_2 )$ is an arbitrary function depending only on $\theta_2$, and
\be
B \equiv 16\,{a^2}\,f(\theta_2)\,
     \left( \cos^{4a}\theta_1 + 
       f(\theta_2)\,
        \sin^{4a} \theta_1 \right).
\ee
Moreover, these solutions must satisfy the K\"ahler condition, $\del_l g_{m\bar{n}} = \del_m g_{l\bar{n}}$ for $l,m,n= w,t$. The K\"ahler condition is satisfied if:
\be
f(\theta_2) = k \cos^{2a-n}\theta_2 \sin^{2a+n}\theta_2,
\ee
where $k, n $ are arbitrary number. In terms of $w,t$ and $z$, these solutions 
become
\ben
g_{m\bar{n}} &=& \del_m G \del_{\bar{n}} \bar{G} + H
\del_m F \del_{\bar{n}} \bar{F}, \label{SOL}\\
H &=& \frac{N \ls^2}{|z|^2 + |F|^2},
\een
where
\ben
F &=&  \sqrt{k} w^{1-p} t^{1+p}/\ls , \\
G &=& \left\{ \ba{cc} \frac{\ls}{2 \sqrt{k}} \log \frac{t}{w}& \mbox{for}\quad p =0,\\
\ls \frac{w^p t^{-p}}{2 p \sqrt{k}} & \mbox{for}\quad p \ne 0.
\ea \right. \label{Gfirst}
\een
Here we define $p \equiv \frac{n}{2a}$. The dilaton $\phi $ is 
$e^{\phi/\sqrt{N} \ls} = H^{\frac{1}{2}}$.

The source equations for these solutions are
\ben
\del_z \del_{\bar{z}} g_{w\bar{w}} + \del_w \del_{\bar{w}} H &=& N(1-p)\ls^2 \delta^{(2)}(z) \delta^{(2)}(w), \\
\del_z \del_{\bar{z}} g_{t\bar{t}} + \del_t \del_{\bar{t}} H &=& N(1+p)\ls^2 \delta^{(2)}(z) \delta^{(2)}(t), \\
\del_z \del_{\bar{z}} g_{w\bar{t}} + \del_w \del_{\bar{t}} H &=& 0.
\een
These equations imply  that the solutions describe intersecting $N(1+p)$ NS5-branes and $N(1-p)$ NS5$'$-branes. Here $N>0$ and $|p|<1$ since both charges of NS5-branes and NS5$'$-branes are positive.
We have fully localized intersecting NS5-branes solutions in the near horizon limit which have 
linear dilaton background (\ref{LDF}). 


\section{M5-brane solutions}\label{secGSOL}
In the previous section, we have assumed the linear dilaton form metric
at first, and then looked for the solutions. In this section, we
consider M5-brane solutions for the equations
(\ref{FSoriginal})-(\ref{SOURCE}) in eleven dimensions and see how the
solutions relate to the linear dilaton metric in Type IIA theory. The
solutions in the near horizon limit are first given in \cite{BFMS} and
we revisit them.

We now  consider the solutions in terms of $v,s,x^{\alpha}$ for $ \alpha =8,9,10$ 
 not fixed coordinates $w,t,y^{\alpha}$ in the decoupling limit. We define that
\be
r^2 \equiv \sum_{\alpha=8}^{10} (x^{\alpha})^2.
\ee
We assume that the determinant $H$ of $g_{m\bar{n}}$ for $v,s$ as
\be
H = \frac{N\,  l_p{}^3}{(r^2 + |F|^2)^{3/2}}, \label{Hfirst}
\ee
where $N$ is a (dimensionless) constant and $F$ is an arbitrary holomorphic function which depends only on $v$ and $s$, $F= F(v,s)$. The determinant $H$ has 
diverges on 
the palace where $r$ and $F$ vanish simultaneously. These singularities specify the 
position of the M5-branes as we will see, therefore, the M5-branes localize at the origin in the overall transverse directions and wrap on the Riemann surface in the $(v,s)$-space. 

The metric elements $g_{m\bar{n}}$ must be a K\"ahler for the M5-brane solution. So we take a K\"ahler potential as $K=K(r,|v|,|s|)$ where we assume that the 
solution has rotational symmetry in the overall transverse directions. 
The sources (\ref{SOURCE})  vanish at the place where M5-branes do not localize. Since $g_{m\bar{n}} = \del_m \del_{\bar{n}} K $, the source equations (\ref{SOURCE}) away from the M5-branes become
\be
\del_m \del_{\bar{n}} \left( \frac{1}{r^2}\del_r (r^2 \del_r K )+ 4 H \right) = 0. \label{SOURCEaway}
\ee
We find the general solution up to some K\"ahler transformation which contain  functions depending only on $r$, 
\be
K = \frac{4 N\,l_p{}^3}{r} \log \left( \sqrt{r^2 + |F|^2} + r \right) + \frac{1}{r} K^{(1)} + K^{(2)}, \label{KP}
\ee
where $K^{(i)}$ for $i=1,2$ are the real functions which do not depend on $r$, so $K^{(i)} = K^{(i)}(|v|,|s|)$.  For simplicity, we take 
\footnote{ The first term in (\ref{KP}) is the same as the solution (12) in \cite{BFMS} up to a K\"ahler transformation. To do this,  we choose $K^{(1)} = -2N\,\log |F|^2$.}
\be
K^{(1)} =0,
\ee
and 
\be
K^{(2)} =  |G|^2,
\ee
where $G$ is a holomorphic function, $G= G(v,s)$. The function $G$ is determined by the condition that the determinant of $g_{m\bar{n}}$ obtained from the K\"ahler potential equals to $H$ again, so
\be
(\det g_{m\bar{n}} =) \del_v \del_{\bar{v}} K \, \del_s \del_{\bar{s}} K - \del_v \del_{\bar{s}} K \, \del_s \del_{\bar{v}} K  = H.
\ee
This equation gives the condition on $G$ that
\be
\del_v F \, \del_s G - \del_s F \, \del_v G = 1. \label{COND}
\ee
Thus, we find the general solution for the M5-branes, which is formally
\be
ds^2 = H^{-\frac{1}{3}} \left( dx_{3+1}^2 + |dG|^2 + H (|dF|^2 + dr^2 +r^2 d\Omega_2{}^2 ) \right), \label{GSOL}
\ee
where $d\Omega_2$ is the standard $S^2$ in the overall transverse directions. 
This metric is naively the same as the parallel M5-branes solution in terms of $F$ and $G$, however, the coordinate map from $(v,s)$ to $(F,G)$ is generally not well-defined on the M5-branes as we saw in the previous section. Therefore, the metric is defined in terms of $v$ and $s$.

The source equations (\ref{SOURCE}) become
\ben
&&\del_{\alpha} \del_{\alpha} g_{m\bar{n}} + 4\del_m \del_{\bar{n}} H  \nonumber \\
&& = \del_m F \del_{n} \bar{F} \left( \del_{\alpha} \del_{\alpha} + 4\del_F \del_{\bar{F}} \right) H \nonumber \\
&& = N\, l_p{}^3 \delta^{(3)} (r) \delta^{(2)}(F) \del_m F \del_{\bar{n}}\bar{F}. \label{SOURCEarb}
\een
So $N$ represents a number of the M5-branes which wrap on the holomorphic curve  $F=0$.

Let us slightly generalize above solutions. We take $H$, instead of
(\ref{Hfirst}), as
\be
H = 1 + \frac{N\, l_p{}^3}{(r^2 + |F|^2)^{3/2}}
\label{Hsecond}
\ee
The source equation away from M5-branes and the
solution of K\"ahler potential for it are the same as (\ref{SOURCEaway})
and (\ref{KP}). We take $K^{(1)}=0$ again and
\be
K^{(2)} = |F|^2 +
|G|^2,
\ee
where $G$ is an arbitrary holomorphic function as like as
before. It is easy to check that the determinant of $g_{m\bar{n}}$ also
gives the same condition as $(\ref{COND})$ on $G$. So the metric and the
source equation are described by the same equations (\ref{GSOL}) and
(\ref{SOURCEarb}) again in which $H$ is given by (\ref{Hsecond}).

The holomorphic function $F$ must have  one dimension  of the length. If we take $F=v$ in particular, $G$ becomes $G=s$. So the solution become the standard 
 parallel $N$ M5-branes solution.

For the general $F$, in the limit $r\to \infty$, the metric becomes asymptotically flat in ${\bf R}^{1,3}$ and the overall transverse space, but,  does not in $(v,s)$-space. So we expect that there are more general solutions which become asymptotically flat far away from the M5-branes.

\subsection{Dimensional reduction to linear dilaton background}
In this subsection, we consider the conditions that the 
solutions (\ref{GSOL}) have the linear dilaton form background in the
decoupling limit.

In the near horizon limit $r$ and $F \to 0$, the first term in H (\ref{Hsecond}) drops out,
\be
H = \frac{N\,  l_p{}^3}{(r^2 + |F|^2)^{3/2}}
\ee
In the decoupling limit, this becomes
\be
H= \frac{1}{l_p{}^6} \tilde{H},
\ee
where we define 
\ben
\tilde{H} &\equiv &\frac{N}{(|y^{\alpha}|^2 + |\tilde{F}|^2)^{\frac{3}{2}}},\\
\tilde{F} &\equiv & F/l_p{}^3.
\een
Then the metric becomes
\be
ds^2 = l_p{}^2 \tilde{H}^{-\frac{1}{3}} \left(
dx_{3+1}^2 + |dG|^2 + \tilde{H} ( |d\tilde{F}|^2 + (dy^{\alpha})^2) \right).
\ee
The condition on $G$ (\ref{COND}) becomes
\be
\del_w \tilde{F} \del_t G - \del_t \tilde{F} \del_w G = 1. \label{COND2}
\ee
Compactifying the metric along the eleventh direction, the metric becomes, in type IIA theory,
\be
ds^2 = dx_{3+1}^2 + |dG|^2 + \tilde{H} ( |d\tilde{F}|^2 + |z|^2). \label{PSOL}
\ee
Here
\be
\tilde{H} = \frac{N\, \ls^2}{ |z|^2 + |\tilde{F}|^2}. \label{PH}
\ee
The dilaton $\phi$ is given by 
\be
e^{\phi/\sqrt{N}\ls} = \tilde{H}^{1/2}.
\ee
From the dimensional analysis, $\tilde{F}$ must have -2 dimensions of the length . We take
\footnote{
In general, we may take 
\be
\tilde{F} = \sum_{i} a_i \, w^{n_i} t^{m_i}\ls^{\frac{n_i + m_i}{2} -2},
\ee
where $n_i$ and $m_i$ are arbitrary real number and $a_i$ are arbitrary complex number.  However, it seems that there is generally no solution for $G$ to the equation (\ref{COND2}) except for the case of the monomial $F$  and the special cases where $F$ reduces to a monomial by a one-to-one coordinate transformation. 
This supports the result of section \ref{secLD} which is general solution for the linear dilaton background.}
\be
\tilde{F} = k \, w^{n} t^{m}\ls^{\frac{n + m}{2} -2}.
\ee
So we consider the intersecting NS5-branes case. 
We choose  the ``radial'' coordinate as
\be
\rho^2 = |z|^2 + |\tilde{F}|^2
\ee
The dilaton $\phi$ depends only on $\rho$ in the six dimensional transverse space $w,t,z$, therefore, $G$ does not depend on $\rho$. From dimensional analysis again, $G$ must have a form as
\footnote{$G$ does not depend on $z$ becouse of the condition (\ref{COND2}) on $G$.}
\be
G = \ls \tilde{G}( w/t ),
\ee
where $\tilde{G}( w/t ) $ does not include $\ls$.
Since the right hand side in (\ref{COND2}) is the constant which does not depend on $\ls$, $F$ must depend only on $\ls^{-1}$. So we have the condition on $n$ and $m$ that
\be
n+ m=2.
\ee
Here we take
\ben
n&=&1-p,\\
m&=&1+p.
\een
The holomorphic function $G$ which determined by (\ref{COND2}) is the same as (\ref{Gfirst}). 
Thus we find that the solution in section \ref{secGSOL} reduces to the
same linear dilaton backgrounds in section \ref{secLD}. There might be
more general solutions to M5-branes as we mentioned, however, if the
solutions reduce to the linear dilaton backgrounds they might become the
form in section \ref{secLD}.

In the parallel NS5-branes case, there is only one (dimensionless) parameter $N$ which is number of the NS5-branes. This corresponds to the size of $S^3$  
surrounding the NS5-branes. On the other hand, 
in the intersecting NS5-branes case, there are three parameter $N,p$ and $k$. Two of them $N$ and $p$ correspond to the number of NS5-branes.  $N$ represents the size of the base $S^3$ in $(z,F(w,t))$-space and $p$ decide the fiber  $G(w,t)$ structure in $(w,t)$-space. The other parameter $k$ corresponds to the size of fiber. In the parallel case, we have the same background structure in the near horizon limit through any pass to close to the NS5-branes. In the intersecting case, however, the background structure depends on the pass in  the near horizon limit, $F(w,t) \to 0$. If we are close to a parallel part of NS5-branes not the intersecting point, the background may be the same as the 
parallel NS5-branes case and the $S^3$ part will be dominant with the small size of fiber $G$.  So if we would like to close to the intersecting point, we need to choose the appreciate pass and this seems to choose the appreciate $k$. In the CFT view, this may correspond to choose the consistent central charge for string theory. 
\section{Relation to  LG models} \label{secLG}
The decoupled non-gravitational theory on the parallel NS5-branes background has studied by using  CFT \cite{ABKS, OV, Kutasov, GKP, GK, EGKRS} in detail. One approach to the theory is holographic description \cite{ABKS, GKP, GK}.  N parallel NS5-branes is given by  the holomorphic curve $v^N=0$ where $v$ is a transverse complex coordinate.  The NS5-branes are the dual to the ALE space \cite{OV, Kutasov},
\be
{\cal F}_2 \equiv v^N + z_1^2 + z_2^2 =0, \label{F2}
\ee
where $z_1$ and $z_2$ are complex coordinates. 
The decoupled theory on the singularity $v=z_1=z_2=0$ relates to string theory on
\be
{\bf R}^{1,5} \times {\bf R}_{\phi} \times S^1 \times LG(W={\cal F}_2), 
\ee
where ${\bf R}_{\phi}$ is labeled by the dilaton $\phi$.

The intersecting NS5-branes solutions (\ref{PSOL}) and (\ref{PH}) are formally the same forms as $N$ parallel NS5-branes solutions in terms of $F$ and $G$. 
$F$ can be considered as transverse complex coordinate like as $v$ in the parallel case.  So we replace $v$ in (\ref{F2}) by $F$ and  the decoupled theory formally relates to string theory on 
\be
{\bf R}^{1,5}\left(={\bf R}^{1,3} \times {\bf C}_G \right) \times {\bf R}_{\phi} \times S^1 \times LG(W={\cal F}_2). \label{PBG}
\ee
Since, however, the coordinate map $w,t \to F,G$ is not well-defined at $w,t=0$ in the intersecting NS5-branes case, $F$ and $G$ are not direct product through $w$ and $t$. 
Therefore the world-volume theory on the intersecting NS5-branes must be described by  string theory on the backgrounds in terms of $w,t$ rather than $F,G$. Replacing $F,G$ with $w,t$, the CFT background (\ref{PBG}) maps to 
\be
{\bf R}^{1,3}\times {\bf R}_{\phi} \times S^1 \times LG(W={\cal F}_3)
\ee
where
\be
{\cal F}_3 = w^{N(1-p)}t^{N(1+p)} + z_1^2 + z_2^2.
\ee
In \cite{GKP},  the authors discussed that, in general,  string theory on the singular $CY_n$ manifold which is described by ${\cal F}_n(z_a)=0$ for $a=0,\ldots  ,n$ relates to string theory on the background,
\be
{\bf R}^{1,9-2n}\times {\bf R}_{\phi} \times S^1 \times LG(W={\cal F}_n).
\ee
Explicit modular invariant partition functions on this background are
constructed in \cite{ES, Mizoguchi, Yamaguchi}.

The duality relation between the intersecting NS5-branes and the conifold 
are discussed from a viewpoint of the SUGRA solutions \cite{DM, KY}. 
The intersecting $N(1-p)$ and $N(1+p)$ NS5-branes are  T-dual to the
generalized conifold ${\cal F}_3 =0$ \cite{AKLM}. 
Therefore the SUGRA solutions (\ref{SOL}) for intersecting NS5-branes are consistent with the results of \cite{GKP}.

\section{Conclusion and discussion}

In this paper, we have found solutions of the near-horizon geometry of
the intersecting NS5-branes, which is a kind of the linear dilation
background. Closed string theory on this background holographically
describes the LST on four dimensional common intersection of branes. We
have been looking for a solution in two different ways. However, both
approaches produce only the near-horizon geometry of a class of the
intersecting branes if we restrict on the linear dilaton background. The
second approach as we mentioned in section \ref{secGSOL} seems to
include more general solutions for the NS5-branes wrapping on the
general Riemann surface. We are very interested in the case that the
Riemann surface is other singular curves as like as Argyres-Dougras
curves \cite{AD}.

We have discussed the relation between our background and the LG
theory. However, explicit connection is not so clear. In the case of
parallel NS5-branes, near-horizon geometry is described by a product of
compact coset spaces on which WZW model is dual to the LG theory. Our
near-horizon solution except for the linear dilaton part might be
non-compact and coset description of this theory is difficult to
understand. So we need the further investigation of the near-horizon
geometry of the NS5-brane wrapping on the $\Sigma$.

Recently, an intersecting brane configuration on which standard model
like theory appears is constructed \cite{AFIRU}. To realize the
Randall-Sundrum scenario on the intersection of branes in superstring
theory, the SUGRA solution for these branes become more important. And
also these solutions are indispensable to understand holographically the
four dimensional QCD or standard model like theories. We hope that our
solution give a hint to solve these problems.

\section*{Acknowledgments}
We would like to thank Hiroyuki Hata and Shin'ichi Imai for very useful
discussions.  Researches of K.O and T.Y are supported in part by the JSPS Research
Fellowships.

\end{document}